\documentclass[review,3p,12pt]{elsarticle}
\usepackage{amsmath}
\usepackage{amsfonts}
\usepackage{amssymb}
\usepackage{graphicx}
\usepackage{dcolumn}
\usepackage{bm}
\usepackage[colorlinks = true, allcolors = blue]{hyperref}
\usepackage{float}
\makeatletter
\def\ps@pprintTitle{%
 \let\@oddhead\@empty
\let\@evenhead\@empty
 \def\@oddfoot{}%
\let\@evenfoot\@oddfoot}
\makeatother

\begin{document}
\begin{frontmatter}
	\title{Modulation of Landau levels and de Haas-van Alphen oscillation in magnetized graphene by uniaxial tensile strain/ stress}
	\author[ad1,ad2]{Dai-Nam~Le\corref{cor1}}
	\ead{ledainam@tdtu.edu.vn} 
		
	\author[ad3]{Van-Hoang~Le}

	\author[ad1,ad2]{Pinaki~Roy\corref{cor1}}
	\ead{pinaki.roy@tdtu.edu.vn}
	
	\cortext[cor1]{corresponding author}	
	
	\address[ad1]{Atomic~Molecular~and~Optical~Physics~Research~Group, Advanced Institute of Materials Science, Ton Duc Thang University, Ho~Chi~Minh~City, Vietnam}
	\address[ad2]{Faculty of Applied Sciences, Ton Duc Thang University, Ho~Chi~Minh~City,~Vietnam}

	\address[ad3]{Department of Physics, Ho Chi Minh City University of Education, 280~An Duong Vuong St.,~Dist. 5,~Ho Chi Minh City,~Vietnam}
	
\begin{abstract}
The strain engineering technique allows us to alter the electronic properties of graphene in various ways. Within the continuum approximation, the influences of strain result in the appearence of a {pseudo-}gauge field 
and modulated Fermi velocity. In this study, we investigate theoretically the effect of linear {uniaxial tensile strain and/or stress}, which makes the Fermi velocity anisotropic, on a magnetized graphene sheet in the presence of an applied electrostatic voltage. More specifically, we analyze the consequences of the anisotropic nature of the Fermi velocity on the structure Landau levels and de Haas - van Alphen (dHvA) quantum oscillation in the magnetized graphene sheet. The effect of the direction of the applied strain has also been discussed. 
\end{abstract}
\end{frontmatter}

\section{\label{sec:intro}Introduction}
Fifteen years ago, the first fabrication of monolayer graphene opened a new era in condensed matter physics as well as materials science \cite{Novoselov2005, Zhang2005}. This was not only the first synthesized atomically thin monolayer but also the first two dimensional Dirac material. Electrons related to $2p_z$ orbital in the honeycomb lattice of graphene surprisingly behave as massless relativistic fermions in $(2+1)$ dimensional space time \cite{Neto2009}. Due to Klein tunneling, only magnetic fields can trap these particles to form the relativistic Landau levels \cite{Goerbig2011, Kuru2009, Masir2011, Roy2012, Downing2016, Le2018a, Le2018b, LE2020113777}. The confining electrons in graphene in magnetic fields open up new possibilities to study some very interesting phenomena such as quantum Hall effect \cite{Bolotin2009, Skachko2009, Son2015}, de Haas - van Alphen (dHvA) oscillation \cite{Sharapov2004, Ma2010, Ma2011, Ali2014} or collapse of Landau levels \cite{Lukose2007, Peres2007, Ghosh2019} etc. The last phenomenon, namely the collapse of Landau levels, takes place when an electrostatic voltage applied to the magnetized graphene sheet reaches a critical value. In contrast to the magnetic field, the in-plane electric field opposes the formation of Landau levels and consequently the Landau levels collapse if the electric field is large enough \cite{Lukose2007, Peres2007, Ghosh2019, Nath2018}. Because of this effect, it is possible to modulate the dHvA oscillation in graphene's magnetization by electric field \cite{Ma2010, Ma2011, Ali2014}.

It is now known that electronic properties of graphene change when it is subjected to mechanical deformation and manipulating the electronic properties in this way is known as the strain engineering technique. Deforming a graphene sheet can produce a pseudoelectromagnetic field \cite{Vozmediano2008, Vozmediano2010, Pereira2009, Low2010, Levy2010, Guinea2010, DeJuan2013}. Interestingly this pseudoelectromagnetic field depends on the valleys the electrons belong to. These fields not only act similarly to the real electromagnetic field, but the pseudoelectromagnetic fields also open a new avenue to control the valley current of the electron in graphene i.e valleytronics  \cite{Vozmediano2008, Vozmediano2010, Pereira2009, Low2010, Levy2010, Guinea2010, DeJuan2013, Etienne2020}. Recently the strain engineering technique has also been used to control the dHvA oscillation besides using an external electric field \cite{Ma2019}. Apart from the induced pseudoelectromagnetic field, another consequence of deformation or strain is the modulation of Fermi velocity. Unlike unstrained graphene with well-known constant Fermi velocity $v_F \approx c / 300$, straining graphene sheet can make Fermi velocity becomes inhomogeneous or anisotropic \cite{Etienne2020, Pellegrino2011, DeJuan2012, Oliva2015, Downing2017, Naumis2017}. Interestingly electrons can be bound in the presence of inhomogeneous magnetic fields and anisotropic Fermi velocity \cite{Concha_2018, Ghosh2017, Phan2020}. In the latter case, the Fermi velocity in zigzag and armchair direction becomes different resulting in the Dirac cones being tilted \cite{Goerbig2011}, and consequently the electronic properties of the graphene sheet under applied external fields depend on the direction of fields and strain. For example, in Refs. \cite{ Ghosh2019, Concha_2018}, the collapse of Landau levels under a crossed electromagnetic field with uniaxial strain has been studied. The critical electric field depends not only on the magnitude but also on the direction of the applied strain. 

In view of the above observations, we feel it would be of interest to examine how {uniaxial tensile strain or stress} affects the Landau levels and also the dHvA oscillation in magnetized graphene. This could provide a mechanical way to modulate magnetic properties of monolayer graphene. To achieve this purpose, we shall consider a deformed graphene sheet under the influence of a perpendicular magnetic field combining with an in-plane electric field. Solving the Dirac-Weyl equation by using the concept of supersymmetric quantum mechanics \cite{Cooper2000}, we shall determine the Landau levels analytically. Following some earlier works \cite{ Sharapov2004, Jauregui1990}, we {calculate} analytically the chemical potential and magnetization of graphene sheet at zero temperature. It {is} shown that the influence of zigzag strain on the quantum oscillation of both the chemical potential and the magnetization is more significant in comparison with the armchair strain when there is an applied electric field. The organization of the paper is as follows: in Section \ref{sec:2} we derive the solutions of the Dirac-Weyl equation i.e. the Landau levels for deformed graphene under crossed electromagnetic fields; in \ref{sec:3} we study collapse of Landau levels; in Section \ref{sec:4} we examine dHvA oscillation of magnetization; in Section \ref{sec:5}, we discuss the case when direction of the applied electric field is changed; finally, Section \ref{conc} is devoted to a conclusion. 

\section{\label{sec:2}Dynamics of electrons in uniaxially deformed graphene sheet under crossed electromagnetic fields}
Consider a graphene sheet with size $L_x \times L_y$ ($L_x, L_y \gg a$, $a \approx 1.42 \text{\AA}$ is lattice constant) being deformed by a {uniaxial tensile strain or stress}
\begin{equation}\label{eqn:strain}
u_x = \epsilon _Z x, \quad u_y = \epsilon _A y,
\end{equation}
where $x$ and $y$ axes are parallel to zigzag {(ZZ)} and armchair {(AC)} directions respectively. The domains of the coordinates are $-L_x / 2 \leq x \leq L_x / 2$ and $-L_y/2 \leq y \leq L_y/2$ while $\epsilon _{Z,A} > 0$ ($\epsilon _{Z,A} < 0$) represents the strength of {tensile strain (or stress)}. The strain tensor is a diagonal one given by \cite{Landau}
\begin{equation}
u_{xx} = \partial _x u_x =  \epsilon _Z, \quad u_{yy} = \partial _y u_y = \epsilon _A, \quad u_{xy} = \frac{1}{2} \left( \partial _x u_y + \partial _y u_x \right) = 0.
\end{equation}
As a result of this deformation, not only the {pseudo-}gauge potential
\begin{equation}\label{eqn:induce-gauge}
\left\{
\begin{array}{l}
A_{x}^{(s)} = \dfrac{\eta \beta \hbar}{2 e a} \left( u_{xx} - u_{yy} \right) = \dfrac{\eta \beta \hbar}{2 e a} \left( \epsilon _Z - \epsilon _A \right), \\
A_{y}^{(s)} = \dfrac{\eta \beta \hbar}{e a} u_{xy} = 0, \\
\phi ^{(s)} = g \left( u_{xx} + u_{yy} \right) = g \left( \epsilon _Z + \epsilon _A \right),
\end{array}
\right. .
\end{equation}
is induced but also the Fermi velocity becomes anisotropic \footnote{We note that in Ref. \cite{Ghosh2019} graphene under uniaxial strain was considered, the strain being applied in a certain direction while the other direction was simultaneously deformed via Possion ratio. Here we have considered a more general strain such that the deformation in zigzag (x) and armchair direcitons are independent. This means changing the zigzag tensile strain does not change the armchair tensile strain. If we put the zigzag and armchair strain tensor as $\epsilon _Z =  - \frac{\beta}{ 1- \beta} \epsilon$ and $\epsilon _A  = \frac{\beta}{1 -\beta} \epsilon \nu$, our Fermi velocity coincides with the one in Ref. \cite{Ghosh2019} for uniaxial zigzag strain. Meanwhile the Poision ratio in our work is now $\nu = - \epsilon _A / \epsilon _Z$.}
\begin{equation}\label{eqn:velocity}
v_{xx} = v_F \left((1+(1-\beta)\epsilon _Z \right), \quad v_{yy} = v_F \left((1+(1-\beta)\epsilon _A \right), \quad v_{xy} = 0.
\end{equation}
Here the Fermi velocity is $v_F \approx 10^6 \text{ m/s}$, $\beta \approx 2 - 3$ is Gr{\"u}neisen parameter, $g \approx 4 - 20 \text{ V}$ is acoustic coupling constant  and $\eta = \pm 1$ is valley index ($K$ or $K^{\prime}$ valley) \cite{Vozmediano2010, Oliva2015}. However, since the induced pseudo-gauge field is space independent, it does not produce pseudo electromagnetic fields. Meanwhile, to confine the electron, it is necessary to apply a constant electromagnetic field such that the magnetic field is perpendicular to the graphene surface while the electric field is along armchair direction as follows:
\begin{equation}
\vec{B} = B \vec{e}_z, \quad \vec{E} = E \vec{e}_y,
\end{equation}
Then corresponding gauge potentials are \footnote{Here we use Landau gauge for convenience. As strain-induced potential $\phi ^{(s)}$ is not valley-dependent, we eliminate its appearance in energy spectrum by setting scalar potential as $(- \phi ^{(s)})$ at $y=0$.} 
\begin{equation}\label{eqn:mag}
A_{x}^{(a)} = - B y, \quad A_{y}^{(a)} = 0, \quad \phi ^{(a)} = - E y - g \left( \epsilon _Z + \epsilon _A \right).
\end{equation}
Therefore, the low excited electrons are governed by the following stationary 2D Dirac-Weyl equation:
\begin{equation}\label{eqn:Dirac}
\left\{ \eta v_{xx} \sigma _x \left( \hat{p}_x + e A_{x}^{(s)} + e A_{x}^{(a)} \right) + v_{yy} \sigma_y \hat{p}_y - e \left( \phi ^{(s)} + \phi ^{(a)} (y) \right) \right\} \Psi(x,y) = \mathcal{E} \Psi (x,y),
\end{equation}
where the linear momentum operators are $\hat{p}_{x,y} = - i \hbar \partial _{x,y}$ and $\sigma _{x,y}$ are Pauli matrices. It can be seen from the above equation that the linear momentum along $x$-axis is conversed, hence, the pseudospinor can be separated as $\Psi (x,y) = \exp{\left(i k_x x\right)} \psi(y)$. Due to the Born-von Karman boundary condition, the wave number $k_x$ must be quantized as $k_x = 2 \pi N_x / L_x$.

Now, by introducing new notations: cyclotron length $l_B = \sqrt{\hbar / eB}$, dimensionless ratio $\alpha = E/(v_F B)$, strain-induced $k_s = \beta ( \epsilon _Z - \epsilon _A) /2 a$ and energy wave-number $\varepsilon = \mathcal{E} / (\hbar v_F) $, Eq. \eqref{eqn:Dirac} can be rewritten as
\begin{eqnarray}\label{eqn:Dirac-2}
&& \left[ i \sigma _y \partial _y -  \dfrac{\eta \left( 1 + (1-\beta)\epsilon _Z \right)}{ 1 + (1-\beta)\epsilon _A} \left(k_x + \eta k_s - \dfrac{y}{l_B^2} \right) \sigma _x \right. \nonumber\\
&& \left. - \dfrac{\alpha y}{\left(1 + (1-\beta)\epsilon _A \right) l_B^2} + \dfrac{\varepsilon}{1 + (1-\beta)\epsilon _A} \right] \psi (y) = 0.
\end{eqnarray}
Rotating the pseudospinor $\psi (y)$ into $\tilde{\psi} (y)$ as \cite{Ghosh2019, Nath2018, Phan2020, Le2019}
\begin{equation}\label{eqn:rotate}
\psi (y) = \exp \left( \dfrac{\sigma_x}{4} \ln \left( \dfrac{\left( 1 + (1-\beta) \epsilon _Z \right) + \eta \alpha}{ \left( 1 + (1-\beta) \epsilon _Z \right) - \eta \alpha}\right) \right) \tilde{\psi} (y),
\end{equation}
Eq. \eqref{eqn:Dirac-2} becomes
\begin{equation}\label{eqn:Dirac-3}
\begin{pmatrix}
0 & \partial _y + W(y) \\
- \partial _y + W(y) & 0
\end{pmatrix} \begin{pmatrix}
\tilde{\psi} (y)_{+} \\
\tilde{\psi} (y)_{-}
\end{pmatrix} = {\tilde\varepsilon} \begin{pmatrix}
\tilde{\psi} (y)_{+} \\
\tilde{\psi} (y)_{-}
\end{pmatrix},
\end{equation}
where
\begin{equation}\label{eqn:W}
W(y) = \dfrac{\sqrt{\left( 1 + (1-\beta)\epsilon _Z \right)^2 - \alpha^2}}{\left( 1 + (1-\beta)\epsilon _A \right) l_B^2} y - \dfrac{\eta \left[ \alpha \varepsilon - \left( 1 + (1-\beta) \epsilon _Z \right)^2 \left( k_x + \eta k_s \right) \right]}{\left( 1 + (1-\beta)\epsilon _A \right) \sqrt{\left( 1 + (1-\beta)\epsilon _Z \right)^2 - \alpha^2}},
\end{equation}
and
\begin{equation}\label{eqn:eprime}
{\tilde\varepsilon} = \dfrac{\left( 1 + (1-\beta) \epsilon _Z \right) \left[ -\varepsilon + \alpha \left( k_x + \eta k_s \right) \right]}{\left( 1 + (1-\beta)\epsilon _A \right) \sqrt{\left( 1 + (1-\beta)\epsilon _Z \right)^2 - \alpha^2}}.
\end{equation}
Eq. \eqref{eqn:Dirac-3} indicates the supersymmetric nature of the problem \cite{Cooper2000}. Following the supersymmetry formalism \cite{Cooper2000}, we act on the left of Eq. \eqref{eqn:Dirac-3} by the operator
\begin{equation*}
\begin{pmatrix}
0 & \partial _y + W(y) \\
- \partial _y + W(y) & 0
\end{pmatrix}. 
\end{equation*}
Then we obtain the following two independent second-order ordinary differential equations:
\begin{equation}\label{eqn:schr}
\left[ - \partial _y^2 + W^2(y) \pm \dfrac{dW}{dy} \right] {\tilde\psi} _{\pm} (y) = (\tilde\varepsilon) ^2 {\tilde\psi} _{\pm} (y).
\end{equation}
These can be regarded as a pair of energy dependent Schr{\"o}dinger equations \cite{YEKKEN2013} corresponding to 1D shifted harmonic oscillators \cite{Cooper2000}
\begin{eqnarray}\label{eqn:V}
W^2(y) \pm \dfrac{dW}{dy} && = \dfrac{\left( 1 + (1-\beta)\epsilon _Z \right)^2 - \alpha^2}{\left( 1 + (1-\beta)^2 \epsilon _A \right) l_B^4} \left\{ y - \dfrac{\eta l_B^2 \left[ \alpha \varepsilon - \left( 1 + (1-\beta)\epsilon _Z \right)^2 \left( k_x + \eta k_s \right) \right]}{ \left( 1 + (1-\beta)\epsilon _Z \right)^2 - \alpha^2} \right\}^2  \nonumber\\
&& \pm \dfrac{\sqrt{\left( 1 + (1-\beta)\epsilon _Z \right)^2 - \alpha^2}}{\left( 1 + (1-\beta) \epsilon _A \right) l_B^2}.
\end{eqnarray}
Hence, from the discrete spectrum of shifted harmonic oscillator, we easily determine the eigenvalue $\varepsilon _{n,\eta} (k_x)$ i.e the Landau levels
\begin{eqnarray}\label{eqn:LLs}
\mathcal{E}_{n,\eta} (k_x) = \dfrac{\hbar E}{B} k_x^{*} + \text{sign}(n) \sqrt{2 |n| e B^{*} \hbar v_F^2}, \quad n = 0, \pm 1, \pm2, \ldots.
\end{eqnarray}
Here the effective $x$-component of momentum $k_x^{*}$ and the effective magnetic field strength $B^{*}$ are both strain-dependent:
\begin{eqnarray}
&& k_x^{*} = k_x +\dfrac{\eta \beta}{2 a} \left( \epsilon _Z - \epsilon _A  \right) \label{eqn:kx-eff}, \\
&& B^{*} = \dfrac{\left(\left( 1 + (1-\beta)\epsilon _Z \right)^2 v_F^2 B^2 - E^2\right)^{3/2} \left( 1 + (1-\beta)\epsilon _A \right) }{\left( 1 + (1-\beta)\epsilon _Z \right)^2 v_F^3 B^2} \label{eqn:B-eff}.
\end{eqnarray}
Noticeably, the spectrum in Eq. (\ref{eqn:LLs}) remains unchanged under the valley-momentum transformation:
\begin{equation*}
\mathcal{E}_{n,-\eta} \left( k_x + \dfrac{2 \eta \beta}{a} (\epsilon _Z - \epsilon _A) \right) = \mathcal{E}_{n,+\eta} \left( k_x \right).  
\end{equation*} 
It may be noted that beside the spin degeneracy $g_{spin} = 2$, the Landau levels are also valley degenerated $g_{valley} = 2$ and the influence of strain-induced vector potential is nothing more than translating $x$-axis momentum, in other words, it is similar to the Aharonov-Bohm effect. Also when there is no deformation $\epsilon _Z = \epsilon _A = 0$, the spectrum $\mathcal{E}_{n,\eta} (k_x)$ coincides to known results in Refs. \cite{Ma2010, Ma2011, Ali2014}.

Since the potential \eqref{eqn:V} is of harmonic oscillator shifted by
\begin{eqnarray}\label{eqn:center}
y_{n,\eta} (k_x) = \dfrac{\eta\hbar}{eB} \left( \dfrac{\text{sign} (n) E B \sqrt{2 |n| e B^{*} v_F^2 / \hbar}}{\left(1 + (1-\beta)\epsilon _Z \right)^2 v_F^2 B^2 - E^2}  - k_x^* \right),
\end{eqnarray}
the Landau levels are centered around $y_{n,\eta} (k_x)$ and consequently, only the ones which satisfy the following condition
\begin{equation*}
- \dfrac{L_y}{2} \leq y_{n,\eta} (k_x) \leq \dfrac{L_y}{2},
\end{equation*}
are localized inside graphene sheet. Explicitly this condition reads
\begin{equation}\label{eqn:condi}
k_{-} \leq k_x \leq k_{+}, \quad k_{\pm} = \dfrac{\text{sign} (n) E B \sqrt{2 |n| e B^{*} v_F^2 / \hbar}}{\left(1 + (1-\beta)\epsilon _Z \right)^2 v_F^2 B^2 - E^2} - \dfrac{\eta \beta}{2 a}\left( \epsilon _Z - \epsilon _A \right) \pm \dfrac{e B L_y}{2 \hbar}.
\end{equation}
From now on we shall consider just one valley, say $\eta = 1$ ($K$ valley), since the results for the other valley would be identical. 
\section{\label{sec:3}Strain effect on Landau levels}
\begin{figure}[H]
\begin{center}
\includegraphics[width = 0.9 \textwidth]{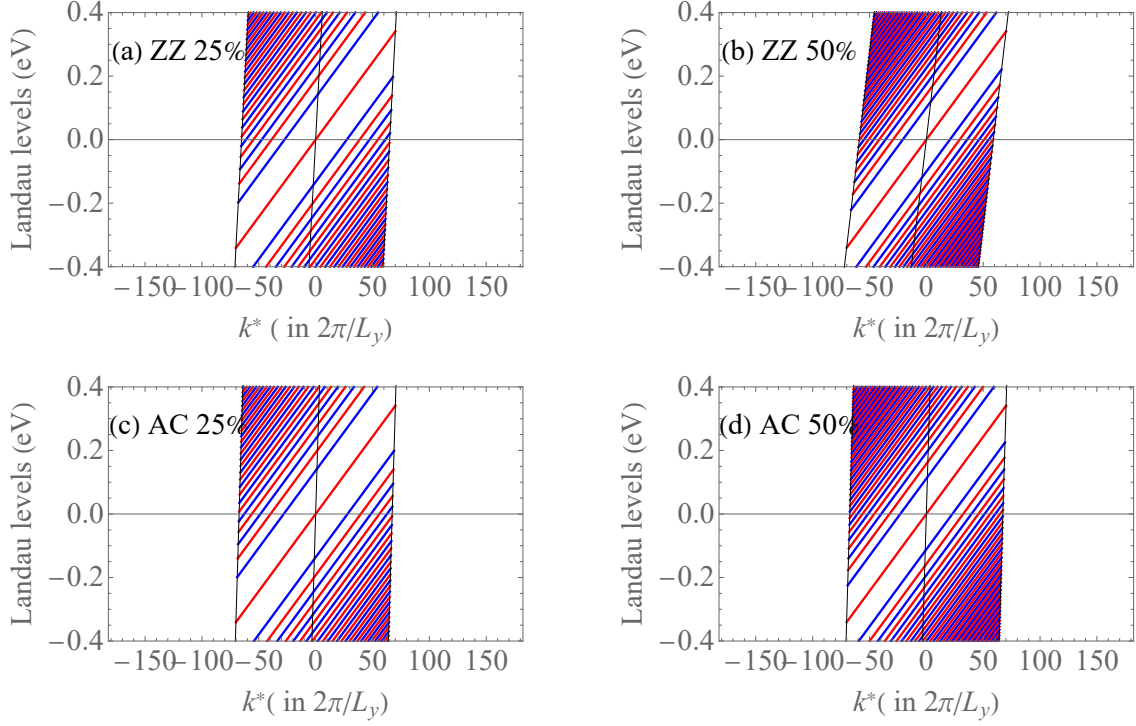}
\caption{\label{fig:1}Landau levels of ${1200 a \times 600 a}$ graphene sheet under different deformations when the electric field is $0.3 \times 10^6 \text{ V/m}$ and magnetic field is $20 \text{ T}$ {i.e the corresponding $\alpha = E / (v_F B) = 0.015$}.  The black lines show the centers of Landau levels. {ZZ and AC are presented for zigzag and armchair deformation respectively while the magnitude of the deformation $\epsilon _{Z,A}$ is expressed by percentage}.}
\end{center}
\end{figure}

For quantitative estimates, we apply a magnetic field of strength $B = 20 \text{ T}$ to the graphene sheet of dimension $1200 a \times 600 a$ and the Gr{\"u}neisen parameter is $\beta = 2$. Fig \ref{fig:1} illustrates Landau levels under different deformations when the electric field is $0.3 \times 10^6 \text{ V/m}$. As can be seen from Fig. \ref{fig:1}, zigzag strain affects the Landau levels more in comparison with the armchair strain of the same magnitude. Not only the Landau levels are titled by the electric field, they are also centered at
\begin{equation}\label{eqn:B2}
\mathcal{E}_{n} = \text{sign}(n) \sqrt{2|n|eB^{**}\hbar v_F^2}, \quad \text{where } B^{**} = \dfrac{\left( 1 + (1-\beta)\epsilon _A \right) \left( 1 + (1-\beta)\epsilon _Z \right)^2 v_F B^2}{ \sqrt{\left( 1 + (1-\beta)\epsilon _Z \right)^2 v_F^2 B^2 - E^2} } ,
\end{equation}
instead of $\text{sign}(n) \sqrt{2|n|eB^{*}\hbar v_F^2}$. {These are the  Landau levels corresponding to the momentum $k^{**} = (k_{+} + k_{-})/2$}.  Comparing to \eqref{eqn:LLs} and \eqref{eqn:center}, the Landau levels can be written as a sum of two parts:
\begin{equation*}
\mathcal{E}_{n, \eta} = \mathcal{E}_{n} + \eta e E y_{n,\eta},
\end{equation*}
where the second part can be interpreted as electric potential at center of the pseudospinor.

Similar to the collapse of Landau levels in unstrained graphene, in the case of strained graphene the Landau levels collapse at a critical electric field depending on strain and is given by 
\begin{equation}\label{eqn:criticalE}
E_c = \left(1+(1-\beta)\epsilon _Z\right) v_F B.
\end{equation}
This formula suggests zigzag strain affects both spacing between the Landau levels and value of the critical electric field strength while armchair strain modulates the spacing only. Fig \ref{fig:2} below illustrates the spacing between first three excited Landau levels (from $n=1$ to $n=3$) and lowest Landau level $n = 0$ near $K/K^{\prime}$ valley depends on electric field strength.
\begin{figure}[H]
\begin{center}
\includegraphics[width = 0.9 \textwidth]{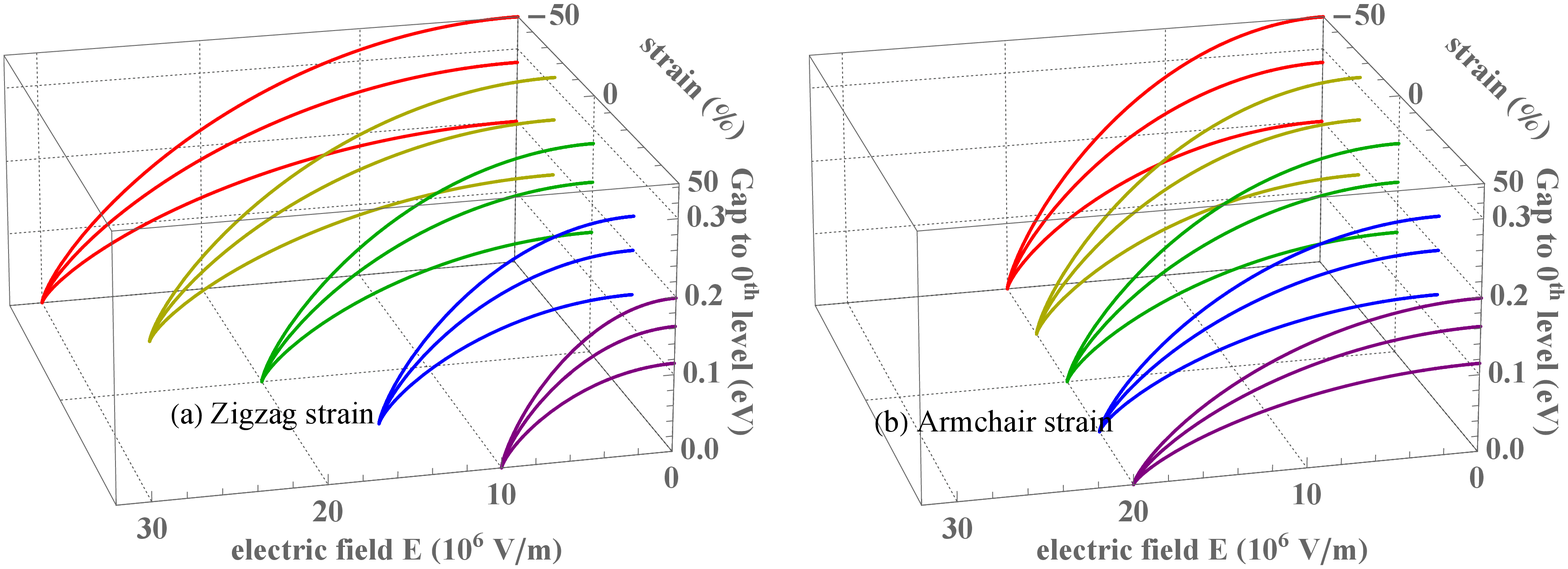}\\
\includegraphics[width = 0.45 \textwidth]{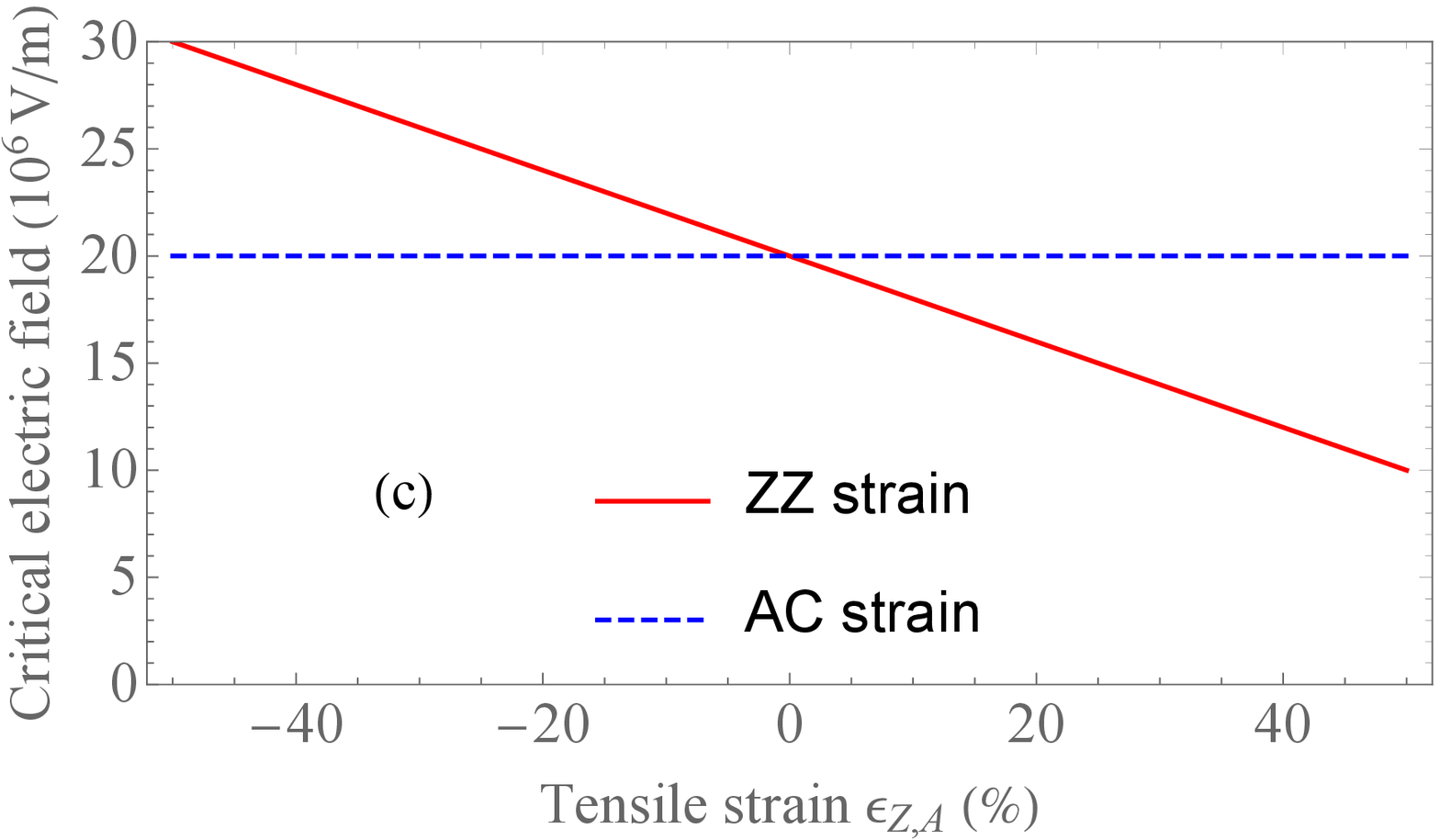}
\caption{\label{fig:2} Modulation of Landau levels by {tensile strain} $\epsilon _{Z,A}$ near $K/K^{\prime}$ valley. (a) Since $\beta \approx 2$, larger {zigzag strain} pulls the critical electric field to the left while larger {zigzag stress} pushes the critical electric field to the right. (b) Armchair strain or stress does not affect on the critical electric field. (c) The critical electric field $E_c$ as function of zigzag (red) and armchair (blue dashed) tensile strain $\epsilon _{Z,A}$.}
\end{center}
\end{figure}

\section{\label{sec:4}Modulation of the magnetization by uniaxial strain}
At zero temperature, all of negative energy levels are fully filled by Dirac sea and graphene exhibits the property of semimetal material. However, when doping electrons into graphene with a concentration $N_0$, the Fermi level i.e chemical potential moves upward to
$\mu _0 = \sqrt{\pi N_0 \hbar ^2 v_F^2}$ and all energy levels between $0$ and $\mu _0$ are filled. Thus graphene now exhibits features of metalic material in which de Haas-van Alphen (dHvA) effect may occur. To investigate the influence of deformation on dHvA effect, we need to determine the magnetization per area $M$ of graphene sheet via the free energy per area $F$:
\begin{equation}\label{eqn:MF}
M = - \dfrac{dF}{dB}, \quad F = \dfrac{1}{L_x L_y} \sum _{\text{filled } n,k_x} \varepsilon _{n,k_x}.
\end{equation}

Taking into account the quantization of the momentum $k_x$ by $\delta k_x = 2 \pi / L_x$ as well as the valley and spin degeneracy, the number of localized Landau states per unit area corresponding to each quantum number $n$ is $D (B) = g_{valley} g_{spin} (k_{+} - k_{-})/\delta k_x = 2 e B /\pi \hbar$. Following Refs. \cite{Ma2010, Ma2011}, we can determine the chemical potential at zero temperature as the magnetic energy $\sqrt{2 n e B^{**} \hbar v_F^2}$ at $n = \left\lfloor N_0/D(B) \right\rfloor + 1$: 
\begin{equation}\label{eqn:mu}
\dfrac{\mu (B, E)}{\mu _0} = \sqrt{\dfrac{\left( 1 + (1-\beta)\epsilon _A \right) \left( 1 + (1-\beta)\epsilon _Z \right)^2 v_F B}{ \sqrt{\left( 1 + (1-\beta)\epsilon _Z \right)^2 v_F^2 B^2 - E^2} } \dfrac{\left\lfloor \nu \right\rfloor + 1}{\nu}} .
\end{equation}
Next we introduce the filling factor $\nu =\mu _0^2/2 e B \hbar v_F^2$ and the floor function $\left\lfloor \nu \right\rfloor$. Subsequently the free energy per unit area $F$ can be found to be
\begin{eqnarray}\label{eqn:F}
F (B, E) = &&  \sqrt{\dfrac{8 e^3 B^2 B^{**} v_F^2}{ \pi^2 \hbar}} \left[ \xi \left(- \dfrac{1}{2} \right) - \xi \left( -\dfrac{1}{2} , \left\lfloor \nu \right\rfloor + 1 \right) + \left( \left\{ \nu \right\} - \dfrac{1}{2} \right) \sqrt{\left\lfloor \nu + 1 \right\rfloor } \right] + \nonumber \\
&& + \left\{ \nu \right\} \left(\left\{ \nu \right\} - 1 \right) \dfrac{e^2 E B L_y}{2 \pi \hbar},
\end{eqnarray}
where the fractional part of $\nu$ is $\left\{ \nu \right\} = \nu - \left\lfloor \nu \right\rfloor$, $\xi (-1/2) \approx -0.207886 $ and $\xi (-1/2, x)$ is Hurwitz zeta function \cite{table2014}. The last term of the free energy comes from the difference of energy due to the electric field when the $(\left\lfloor \nu \right\rfloor + 1)-$th Landau level is partially filled. As can be seen from the expression of the chemical potential and free energy, the ceiling or floor functions are piecewise function and only continuously vary when the filling factor is between two integers $n \leq \nu < n+1$. Thus the discontinuous profile of the free energy can be seen with the period $\Delta (1/B) = 2 e \hbar v_F^2 / \mu_0^2$ corresponds to $\Delta \nu = 1$. From Eqs. \eqref{eqn:MF} and \eqref{eqn:F}, the magnetization per unit area at zero temperature can be determined as
\begin{eqnarray}\label{eqn:M}
M (B, E) = && - \dfrac{3 \left( 1 + (1-\beta)\epsilon _Z \right)^2 v_F^2 B^2 - 4 E^2}{ 2 \left[\left( 1 + (1-\beta)\epsilon _Z \right)^2 v_F^2 B^2 - E^2\right]} \times \nonumber\\
&& \times \sqrt{\dfrac{8 e^3 B^{**} v_F^2}{ \pi^2 \hbar}}  \left[ \xi \left(- \dfrac{1}{2} \right) - \xi \left( -\dfrac{1}{2} , \left\lfloor \nu \right\rfloor + 1 \right) + \left( \left\{ \nu \right\} - \dfrac{1}{2} \right) \sqrt{\left\lfloor \nu \right\rfloor + 1} \right] + \nonumber\\
&& - \left(1 - \delta \left(\left\{\nu \right\}\right) \right) \sqrt{\dfrac{8 e^3 B^{**} v_F^2 \nu ^2 \left( \left\lfloor \nu \right\rfloor + 1 \right)}{ \pi^2 \hbar}} - \left\{ \nu \right\} \left(\left\{ \nu \right\} - 1 \right) \dfrac{e^2 E L_y}{2 \pi \hbar} + \nonumber\\
&& + \left(1 - \delta \left(\left\{\nu \right\}\right) \right) \nu \left( \left\{ \nu \right\} - \dfrac{1}{2} \right) \dfrac{e^2 E L_y}{\pi \hbar}.
\end{eqnarray}
The Dirac $\delta$ function arises from the derivative of the fractional part $\left\{ \nu \right\}$ which is undetermined i.e. it does not exist at integer value of filling factor $\nu$.

Let us now consider a graphene sheet as in Section \ref{sec:3} when it is negatively doped by electrons with a concentration $N_0 \approx 290 \; \mu \text{m}^2$ i.e. initial chemical potential is $\mu _0 \approx 20 \text{ meV}$. Fig \ref{fig:3} below show how chemical potential and magnetization curves are modulated by zigzag and armchair deformations.
\begin{figure}[H]
\begin{center}
\includegraphics[width = 0.9 \textwidth]{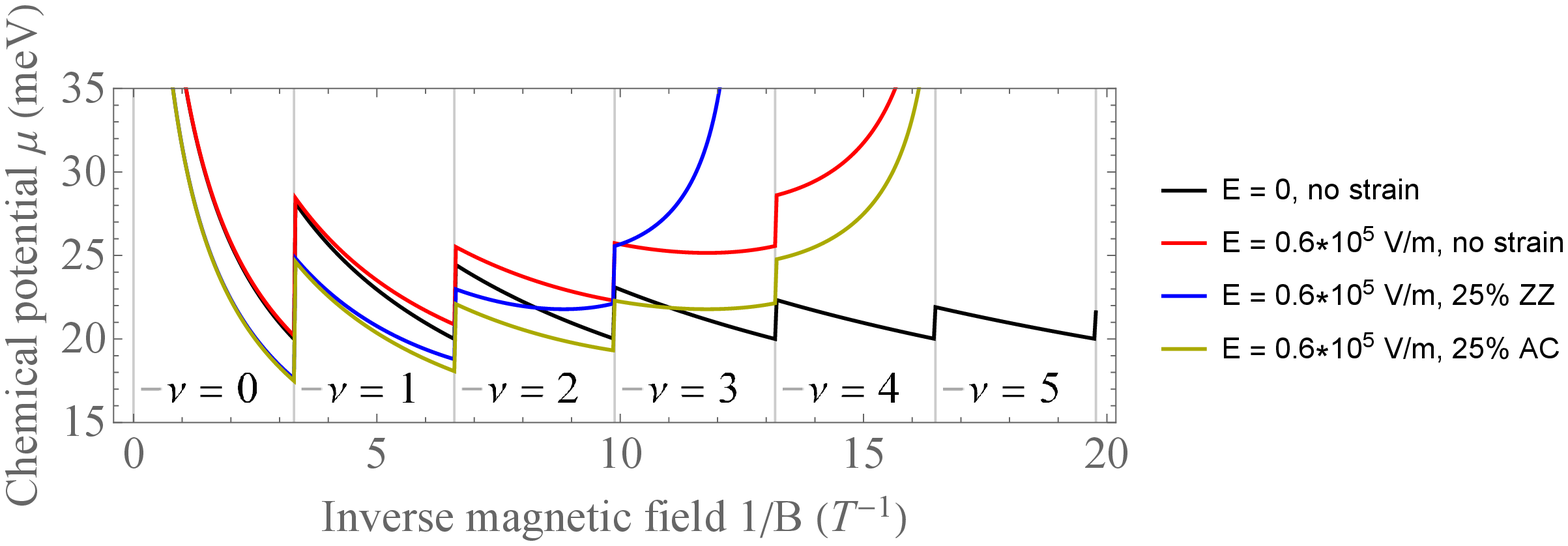}\\
\includegraphics[width = 0.94 \textwidth]{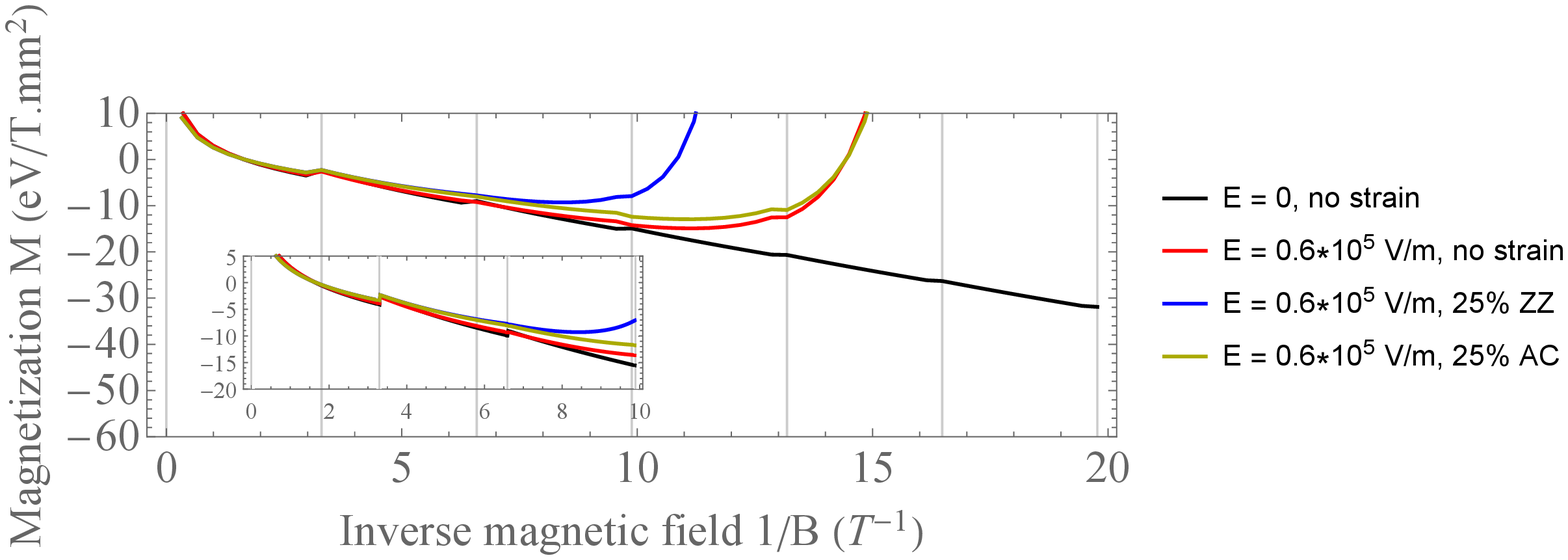}
\caption{\label{fig:3} Chemical potential $\mu$ (above) and magnetization $M$ per unit area (below) versus inverse magnetic field strength $1/B$ under different configurations of applied electric field and strain: (black) no electric field and no strain, (red) $E = 0.6 \times 10^5 \text{ V/m}$ with no strain, (blue) $E = 0.6 \times 10^5 \text{ V/m}$ with $25\%$ zigzag {strain} and (yellow) $E = 0.6 \times 10^5 \text{ V/m}$ with $25\%$ armchair {strain}. Grid lines correspond to integer filling factor $\nu = 0$ to $\nu = 5$.}
\end{center}
\end{figure}

From Section \ref{sec:3} it follows that if there is applied electrostatic voltage, Landau levels only exists when $E \leq \left(1+(1-\beta)\epsilon _Z\right) v_F B$. Also the chemical potential as well as magnetization per unit area oscillates with the period of $\Delta (1/B) = 2 e \hbar v_F^2 / \mu_0^2$ {which surprisingly solely depends on the magnetic field strength $B$ and does not depend on neither type of strain}. Hence number of dHvA oscillation period is only $ \nu _{max} = \left(1+(1-\beta)\epsilon _Z\right) \mu _0 / 2 e \hbar v_F E$. Increasing the electric field will reduce this number of dHvA oscillation period. Noticeably the number of dHvA oscillation period is only controlled by zigzag strain. Fig. \ref{fig:3} shows this effect. The armchair strain only influences the magnitude of both the chemical potential and magnetization. This is a direct consequence from the fact that armchair strain only change the spacing between Landau levels but cannot induce a collapse as zigzag strain. The amplitude of oscillation is however quite small making it difficult to observe dHvA effect, thus it would be useful to separate the total magnetization \eqref{eqn:M} into three parts: regular term $M_{reg}$, oscillating term $M_{osc}$ and electric term $M_{elec}$. Since the oscillation arises from appearance of floor function of filling factor $\left\lfloor \nu \right\rfloor$, the regular magnetization and oscillating magnetization read
\begin{eqnarray}\label{eqn:Mreg}
M _{reg}(B, E) = - \dfrac{3 \left( 1 + (1-\beta)\epsilon _Z \right)^2 v_F^2 B^2 - 4 E^2}{ 2 \left[\left( 1 + (1-\beta)\epsilon _Z \right)^2 v_F^2 B^2 - E^2\right]} \sqrt{\dfrac{8 e^3 B^{**} v_F^2}{ \pi^2 \hbar}} \left[ \xi \left(- \dfrac{1}{2} \right) + \dfrac{2 \nu ^{3/2}}{3} \right],
\end{eqnarray}
and
\begin{eqnarray}\label{eqn:Mosc}
M_{osc} (B, E) = && - \dfrac{3 \left( 1 + (1-\beta)\epsilon _Z \right)^2 v_F^2 B^2 - 4 E^2}{ 2 \left[\left( 1 + (1-\beta)\epsilon _Z \right)^2 v_F^2 B^2 - E^2\right]} \times \nonumber\\
&& \times \sqrt{\dfrac{8 e^3 B^{**} v_F^2}{ \pi^2 \hbar}}  \left[ - \dfrac{2 \nu ^{3/2}}{3} - \xi \left( -\dfrac{1}{2} , \left\lfloor \nu \right\rfloor + 1 \right) + \left( \left\{ \nu \right\} - \dfrac{1}{2} \right) \sqrt{\left\lfloor \nu \right\rfloor + 1} \right] + \nonumber\\
&& - \left(1 - \delta \left(\left\{\nu \right\}\right) \right) \sqrt{\dfrac{8 e^3 B^{**} v_F^2 \nu ^2 \left( \left\lfloor \nu \right\rfloor + 1 \right) }{ \pi^2 \hbar}} .
\end{eqnarray}
The third term of magnetization due to the electric field arises when the $(\left\lfloor \nu \right\rfloor + 1)$th Landau level is partially filled and it vanishes when there is no electric field:
\begin{eqnarray}\label{eqn:Melec}
M _{elec}(B, E) = - \left\{ \nu \right\} \left(\left\{ \nu \right\} - 1 \right) \dfrac{e^2 E L_y}{2 \pi \hbar} + \left(1 - \delta \left(\left\{\nu \right\}\right) \right) \nu \left( \left\{ \nu \right\} - \dfrac{1}{2} \right) \dfrac{e^2 E L_y}{\pi \hbar}.
\end{eqnarray}
Fig \ref{fig:4} below shows how electric field and strain can influence each term of magnetization. Electric field creates not only the $M_{elec}$ term and also reduces the oscillation in $M_{osc}$. Both zigzag and armchair strains do not affect $M_{elec}$. However, only zigzag strain significantly modulates the oscillating term of magnetization while the armchair strain do not. Since the magnitude of the oscillating term is dominated by the regular term, it is difficult to observe the dHvA effect in the total magnetization (see in Fig. \ref{fig:3}). {Note that, again, the period of quantum oscillations cannot be controlled by either the zigzag or the armchair strains because it only comes from the degeneracy of the Landau levels}. 

\begin{figure}[H]
\begin{center}
\includegraphics[width = 0.9 \textwidth]{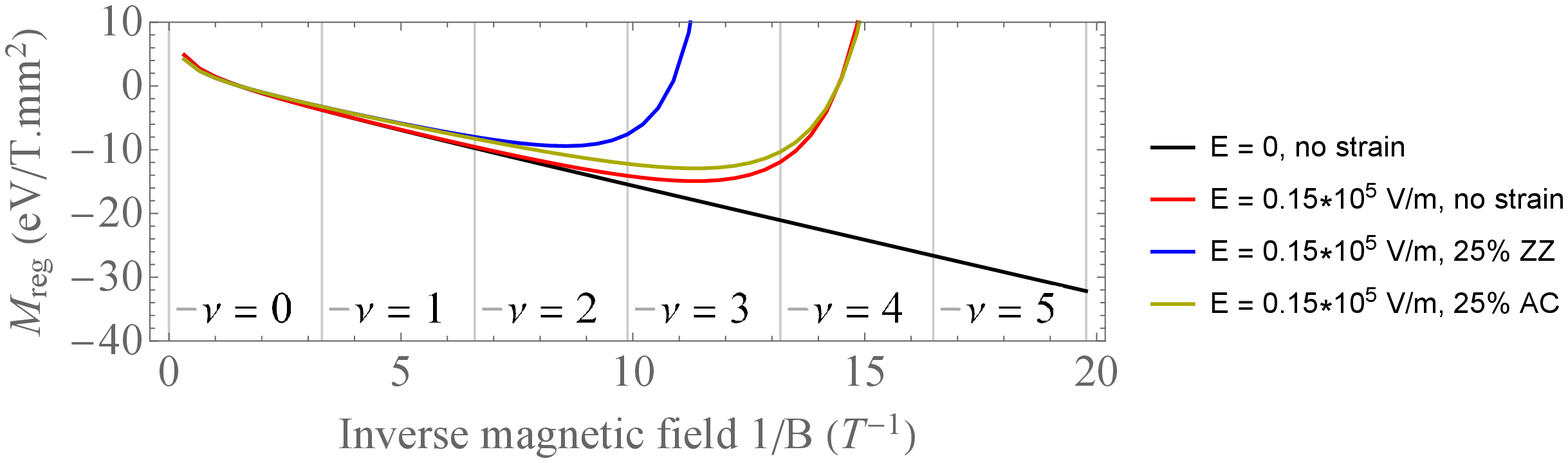}\\
\includegraphics[width = 0.9 \textwidth]{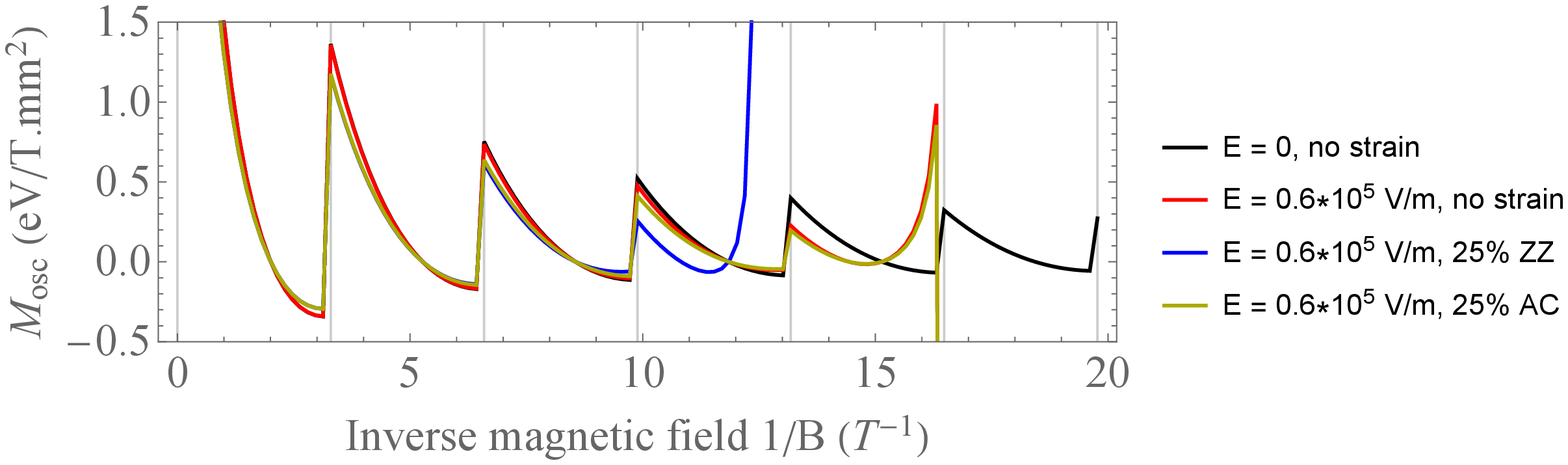}\\
\includegraphics[width = 0.9 \textwidth]{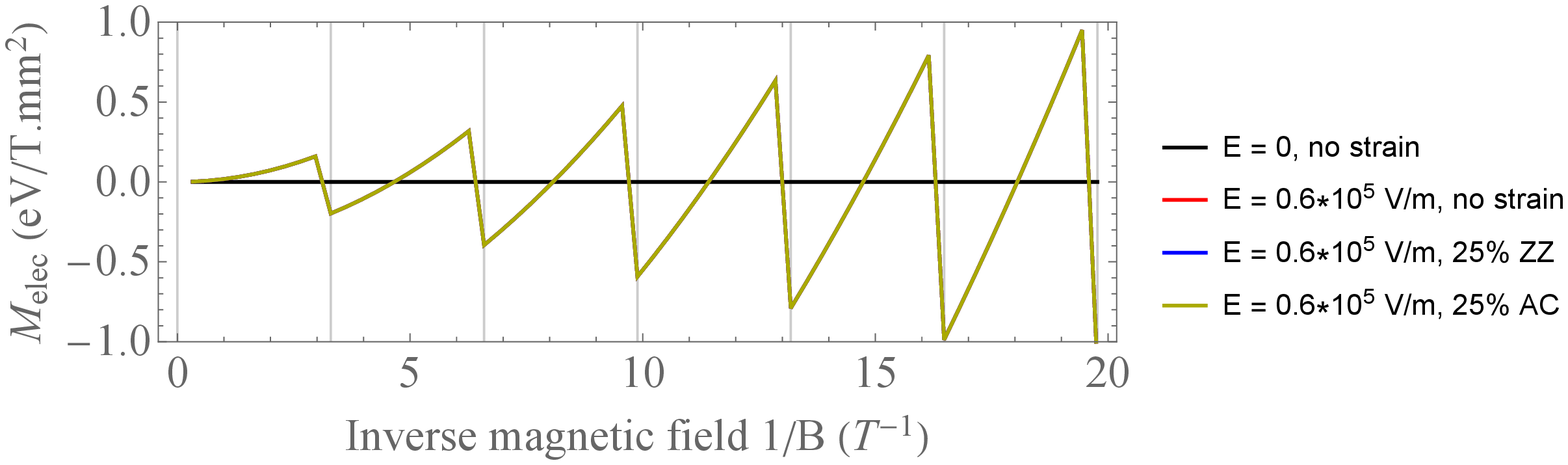}
\caption{\label{fig:4} Each term of magnetization ($M_{reg}, M_{osc}, M_{elec}$)  versus inverse magnetic field strength $1/B$ under different configurations of applied electric field and strain: (black) no electric field and no strain, (red) $E = 0.6 \times 10^5 \text{ V/m}$ with no strain, (blue) $E = 0.6 \times 10^5 \text{ V/m}$ with $25\%$ zigzag {strain} and (yellow) $E = 0.6 \times 10^5 \text{ V/m}$ with $25\%$ armchair {strain}. Grid lines correspond to integer filling factor $\nu = 0$ to $\nu = 5$.}
\end{center}
\end{figure}

{Also, since the regular component of our system mostly decreases (so long  as the electric field is lower than the critical one) versus $1/B$, we may consider to measure the differential magnetic susceptibility
$\chi = \frac{\partial M}{\partial H} \approx \mu_0 \frac{\partial M}{\partial B} $
in the context of experiements. Then the regular term could be
$\chi _{reg} = \mu _0 \frac{\partial M}{\partial B} = - \frac{ \mu_0}{B^2} \frac{\partial M}{\partial (1/B)}$
in which the last factor $\frac{\partial M}{\partial (1/B)}$ slightly varies i.e $ B^2 \chi _{reg}$ is almost a constant. Hence only the oscillation part $B^2 \chi _{osc}$ and the electric part $B^2 \chi _{elec}$ mainly contribute to the non constant profile of $B^2 \chi$.}

\section{\label{sec:5}Discussion on changing direction of electric field}

Finally, we would like to note that if the electric field is along the zigzag direction, the gauge potential should be of the form $A_x^{(a)} = 0, A_y^{(a)} = Bx$ and $\phi ^{(a)} = Ex$. Then the resulting Hamiltonian, at the first sight, is not as the same as in Eq. \eqref{eqn:Dirac}:
\begin{equation}
	\hat{H} =  \left\{ \eta v_{xx} \sigma _x ( \hat{p}_x + e A^{(s)}_{x} ) + v_{yy} \sigma _{y} (k + Bx) - e E x \right\}.
\end{equation}
The cause of this apparently different form is due to the combination of  strain-induced gauge potential $A_x^{(s)}$ and the momentum $\hat{p}_x$ to from a covariant derivative
\begin{equation*}
	\hat{p}_x + e A^{(s)}_{x} = - i \hbar \left( \partial _x + i  \dfrac{e A^{(s)}_{x}}{\hbar} \right).
\end{equation*}

Fortunately, we can remove the strain-induced gauge potential $A_x^{(s)}$ by using a gauge transformation \cite{peskin}. Explicitly, the pseudospinor $\Psi (x,y)$ should be transformed as  
\begin{equation}\label{psi5}
	\Psi (x,y) =  \exp{\left( - \dfrac{i e U(x)}{ \hbar} \right) } \Psi ^{\prime} (x,y).
\end{equation}
When the covariant derivative $\hat{p}_x + e A^{(s)}_{x}$ is applied on Eq. (\ref{psi5}), we obtain
\begin{equation}
	\left(\hat{p}_x + e A^{(s)}_{x} \right) \Psi (x,y) = \exp{\left( - \dfrac{i e U(x)}{ \hbar} \right) } \left(\hat{p}_x + e A^{(s)}_{x} - e \dfrac{dU(x)}{dx} \right) \Psi ^{\prime} (x,y) .
\end{equation} 
Thus $A_x^{(s)}$ can be eliminated when
\begin{equation}
	\dfrac{dU(x)}{dx} =  A^{(s)}_{x} \Rightarrow U(x) = \int A^{(s)}_{x} dx = A^{(s)}_{x} x .
\end{equation}

Now, applying the gauge transformation found above, namely, $\Psi (x,y) =  \exp{\left( - \dfrac{\imath e A^{(s)}_{x} x}{ \hbar} \right) } \Psi ^{\prime} (x,y)$, the Dirac-Weyl equation turns to
\begin{equation}
	\hat{H}^{\prime} \Psi ^{\prime} (x,y) = \mathcal{E} \Psi ^{\prime} (x,y),
\end{equation}
where the transformed Hamiltonian $\hat{H}^{\prime}$ is given as
\begin{eqnarray}
	\hat{H}^{\prime}  && = \exp \left( \imath \frac{e A^{s}_{x} x}{\hbar} \right) \hat{H} \exp \left( - \imath \frac{e A^{s}_{x} x}{\hbar} \right) \nonumber\\
	&& = \eta v_{xx} \sigma _x \hat{p}_x + v_{yy} \sigma _{y} (k + Bx) - e E x  .
\end{eqnarray}
Thus the gauge transformed Hamiltonian $\hat{H}^{\prime}$ is identical to the one in Eq. \eqref{eqn:Dirac} if $x$ is changed to $y$ and vice versa. Meanwhile the zigzag strain and the armchair strain would interchange their roles in dHvA oscillation of magnetization.
\section{\label{conc}Conclusion}
In summary, we have examined the effect of linear {strain or stress} on both the collapse of Landau levels and de Haas-van Alphen oscillation of magnetization in a magnetized graphene sheet in the presence of an electric field. The linear uniaxial strain makes Fermi velocity anisotropic, thus observation of physical quantities depends on the type of deformation or strain. As it has been shown here, the influence of zigzag strain is more significant than armchair strain because the former governs both the magnitude and the critical electric field strength for the Landau levels to exist while the latter only affects the {spacing between} the Landau levels. Consequently, only zigzag strain can control the number of oscillations in the magnetization of the graphene sheet. However, both kinds of strain cannot change the period of quantum oscillation in magnetization since it comes from the degeneracy of the Landau levels, which does not depend on strain. 


\bibliographystyle{elsarticle-num} \bibliography{anisotropic}
\end{document}